\documentclass[12pt]{article}
\usepackage{amssymb}
\usepackage{rotating}

\def\hybrid{\topmargin 0pt      \oddsidemargin 0pt
        \headheight 0pt \headsep 0pt
        \voffset=-0.5cm
        \hoffset=-0.25in
        \textwidth 6.75in
        \textheight 9.5in       
        \marginparwidth 0.0in
        \parskip 5pt plus 1pt   \jot = 1.5ex}
\catcode`\@=11
\def\marginnote#1{}

\newcount\hour
\newcount\minute
\newtoks\amorpm
\hour=\time\divide\hour by60 \minute=\time{\multiply\hour by60 \global\advance\minute by-\hour}
\edef\standardtime{{\ifnum\hour<12 \global\amorpm={am}%
        \else\global\amorpm={pm}\advance\hour by-12 \fi
        \ifnum\hour=0 \hour=12 \fi
        \number\hour:\ifnum\minute<10 0\fi\number\minute\the\amorpm}}
\edef\militarytime{\number\hour:\ifnum\minute<10 0\fi\number\minute}

\def\draftlabel#1{{\@bsphack\if@filesw {\let\thepage\relax
   \xdef\@gtempa{\write\@auxout{\string
      \newlabel{#1}{{\@currentlabel}{\thepage}}}}}\@gtempa
   \if@nobreak \ifvmode\nobreak\fi\fi\fi\@esphack}
        \gdef\@eqnlabel{#1}}
\def\@eqnlabel{}
\def\@vacuum{}
\def\draftmarginnote#1{\marginpar{\raggedright\scriptsize\tt#1}}
\def\draftlabel#1{{\@bsphack\if@filesw {\let\thepage\relax
   \xdef\@gtempa{\write\@auxout{\string
      \newlabel{#1}{{\@currentlabel}{\thepage}}}}}\@gtempa
   \if@nobreak \ifvmode\nobreak\fi\fi\fi\@esphack}
        \gdef\@eqnlabel{#1}}
\def\@eqnlabel{}
\def\@vacuum{}
\def\draftmarginnote#1{\marginpar{\raggedright\scriptsize\tt#1}}

\def\draft{\oddsidemargin -.5truein
        \def\@oddfoot{\sl preliminary draft \hfil
        \rm\thepage\hfil\sl\today\quad\militarytime}
        \let\@evenfoot\@oddfoot \overfullrule 3pt
        \let\label=\draftlabel
        \let\marginnote=\draftmarginnote
   \def\@eqnnum{(\theequation)\rlap{\kern\marginparsep\tt\@eqnlabel}%
\global\let\@eqnlabel\@vacuum}  }

\def\numberbysection{\@addtoreset{equation}{section}
        \def\theequation{\thesection.\arabic{equation}}}

\def\underline#1{\relax\ifmmode\@@underline#1\else
        $\@@underline{\hbox{#1}}$\relax\fi}

\def\titlepage{\@restonecolfalse\if@twocolumn\@restonecoltrue\onecolumn
     \else \newpage \fi \thispagestyle{empty}\c@page\z@
        \def\thefootnote{\fnsymbol{footnote}} }

\def\endtitlepage{\if@restonecol\twocolumn \else  \fi
        \def\thefootnote{\arabic{footnote}}
        \setcounter{footnote}{0}}  

\numberbysection \hybrid

\newcounter{mo}

\newcommand{\tr}{{\rm tr}}
\newcommand{\ti}[1]{\tilde{#1}}

\newcommand{\la}{\lambda}

\newcommand{\al}{\alpha}

\newcommand{\bfq}{{\bf{q}}}

\def\beq{\begin{equation}}
\def\eq{\end{equation}}
\def\p{\partial}

\newcommand{\mat}[4]{\left(\begin{array}{cc}{#1}&{#2}\\ \ \\{#3}&{#4}
\end{array}\right)}

\newcommand{\mats}[4]{\left(\begin{array}{cc}{#1}&{#2}\\ {#3}&{#4}
\end{array}\right)}

\def\res{\mathop{\hbox{Res}}\limits}

\begin{document}

\setcounter{page}{1}

\date{}
\date{}
\vspace{50mm}

\begin{flushright}
 ITEP-TH-18/14\\
\end{flushright}
\vspace{0mm}

\begin{center}
\vspace{0mm}
{\LARGE{Classical integrable systems and soliton equations}}
 \\ \vspace{2mm} {\LARGE{related to eleven-vertex
R-matrix} }
\\
\vspace{12mm} {\large \ \ \ \ \ {A. Levin}\,$^{\flat\,\sharp}$ \ \ \
\ {M. Olshanetsky}\,$^{\sharp\,\natural}$
 \ \ \ {A. Zotov}\,$^{\diamondsuit\, \sharp\, \natural}$ }\\
 \vspace{10mm}

 \vspace{2mm} $^\flat$ -- {\small{\sf 
 NRU HSE, Department of Mathematics,
 Myasnitskaya str. 20,  Moscow,  101000,  Russia}}\\
 \vspace{2mm} $^\sharp$ -- {\small{\sf 
 ITEP, B. Cheremushkinskaya str. 25,  Moscow, 117218, Russia}}\\
 \vspace{2mm} $^\natural$ -- {\small{\sf MIPT, Inststitutskii per.  9, Dolgoprudny,
 Moscow region, 141700, Russia}}\\
\vspace{2mm} $^\diamondsuit$ -- {\small{\sf Steklov Mathematical
Institute  RAS, Gubkina str. 8, Moscow, 119991,  Russia}}\\
\end{center}

\begin{center}\footnotesize{{\rm E-mails:}{\rm\ \
 alevin@hse.ru,\  olshanet@itep.ru,\  zotov@mi.ras.ru}}\end{center}

 \begin{abstract}
In our recent paper we suggested a natural construction of the
classical relativistic integrable tops in terms of the quantum
$R$-matrices. Here we study the simplest case -- the 11-vertex
$R$-matrix and related ${\rm gl}_2$ rational models. The
corresponding top is equivalent to the 2-body Ruijsenaars-Schneider
(RS) or the 2-body Calogero-Moser (CM) model depending on its
description. We give different descriptions of the integrable tops
and use them as building blocks for construction of more complicated
integrable systems such as Gaudin models and classical spin chains
(periodic and with boundaries). The known relation between the top
and CM (or RS) models allows to re-write the Gaudin models (or the
spin chains) in the canonical variables. Then they assume the form
of $n$-particle integrable systems with $2n$ constants. We also
describe the generalization of the top to 1+1 field theories. It
allows us to get the Landau-Lifshitz type equation. The latter can
be treated as non-trivial deformation of the classical continuous
Heisenberg model. In a similar way the deformation of the principal
chiral model is also described.
 \end{abstract}

\newpage

{\small{

\tableofcontents

}}

\section{Introduction}
\setcounter{equation}{0}

In this paper we deal with the quantum 11-vertex $R$-matrix
\cite{Cherednik}\footnote{See also \cite{Smirnov}, where it was
derived by non-trivial limiting procedure from the Baxter quantum
elliptic $R$-matrix \cite{Baxter}.}:
 \beq\label{w001}
 \begin{array}{c}
  \displaystyle{R^\hbar(z)=
 \left( \begin{array}{cccc} {\hbar}^{-1}+{z}^{-1}&0&0&0
\\\noalign{\medskip}-\hbar-z&{\hbar}^{-1}&{z}^{-1}&0\\\noalign{\medskip}
-\hbar-z&{z}^{-1}&{\hbar}^{-1}&0\\\noalign{\medskip}-{\hbar}^{3}-2\,z{\hbar}^{2}-2\,\hbar
\,{z}^{2}-{z}^{3}&\hbar+z&\hbar+z&{\hbar}^{-1}+{z}^{-1}
\end{array} \right)
  }
 \end{array}
 \eq


\noindent{\bf Relativistic integrable tops.} It was recently
observed \cite{LOZ7} that (\ref{w001}) is the simplest example of
the quantum rational ${\rm gl}_N$ $R$-matrix appearing from the
classical relativistic integrable top.  The relativistic top is
defined by its classical Lax operator in terms of the quantum
$R$-matrix\footnote{Notice that the Planck constant in the
$R$-matrix is replaced by the relativistic deformation parameter
$\eta$ of the Ruijsenaars-Schneider (RS) type \cite{Ruijs1}.
 In
the RS model it equals to the ratio of the coupling constant to the
light speed.}
 \beq\label{ww003}
  \begin{array}{c}
  \displaystyle{
 { L^\eta}(z,\mathcal S)\equiv L^\eta(z)=\tr_2 \left(R^{\,\eta}_{12}(z) \mathcal S_2\right)\,,\
 \ \mathcal S=\res\limits_{z=0}  L^\eta(z)\,,\ \ \mathcal S_2=1\otimes \mathcal S
 }
 \end{array}
 \eq
with the spectral parameter $z$. The dynamical variables are the
components of $2\times 2$ matrix $\mathcal S\in {\rm gl}_2$, which
is the residue of $L^\eta(z,\mathcal S)$. The Poisson structure is
generated by the quadratic $r$-matrix structure
 \beq\label{ww004}
 \begin{array}{c}
  \displaystyle{
\{L^\eta_1(z)\,, L^\eta_2(w)\}=[ L^\eta_1(z)\,
L^\eta_2(w),r_{12}(z-w)]\,,
 }
 \end{array}
 \eq
where $r_{12}(z-w)$ is the classical $r$-matrix (the classical limit
of (\ref{w001}))
\footnote{The non-relativistic rational ${\rm gl}_N$
tops were described in \cite{AASZ}, while the ${\rm sl}_2$ case was
derived previously in \cite{Burban1}.}:
 \beq\label{w005}
   \displaystyle{
{r}_{12}(z)=
 \left(\begin{array}{cccc}
1/z & 0 & 0 & 0\\ -z & 0 & 1/z & 0\\ -z & 1/z & 0 & 0\\ -z^3 & z & z
& 1/z
 \end{array}
 \right)
 }
 \eq
 The equations of motion are of the Euler type:
  \beq\label{w00522}
   \displaystyle{
\dot{\mathcal S}=[{\mathcal S}, J^\eta({\mathcal S})]\,,
 }
 \eq
where the inverse inertia tensor $J^\eta$ is given by (\ref{we07}),
(\ref{w41}).

In the non-relativistic limit $\eta\to 0$ the $r$-matrix structure
(\ref{ww004}) becomes the linear:
 \beq\label{ww006}
 \begin{array}{c}
  \displaystyle{
\{L_1(z)\,, L_2(w)\}=[ L_1(z) + L_2(w),r_{12}(z-w)]\,.
 }
 \end{array}
 \eq
Similarly to (\ref{ww003}) the Lax matrix is expressed in terms of
the classical $r$-matrix
 \beq\label{ww007}
  \begin{array}{c}
  \displaystyle{
 { L}(z,S)\equiv L(z)=\tr_2 \left(r_{12}(z) S_2\right)\,,\
 \  S=\res\limits_{z=0}  L(z)\in {\rm gl}_2\,.
 }
 \end{array}
 \eq
It leads to the following equations:
 \beq\label{ww0071}
  \begin{array}{c}
  \displaystyle{
\dot S=[S,J(S)]\,.
 }
 \end{array}
 \eq
While the Lax matrix (\ref{ww003}) is the quasi-classical limit of
the quantum $L$-operator, the standard description of the classical
quadratic Poisson structures deals with the different Lax matrix:
 \beq\label{ww0081}
 \begin{array}{c}
  \displaystyle{
{\tilde L}(z,\tilde S)={\tilde S}_0\, 1_{2\times 2}+L(z,\tilde
S)-\frac{1}{2}\,\tr L(z,\tilde S)\,1_{2\times 2}\,,\ \ \tilde
S_0=\tr\,\tilde S/2\,,\ \  \ti S\in {\rm gl}_2\,.
  }
 \end{array}
 \eq
The latter is independent of $\eta$. Similarly to the
non-relativistic case it is defined in terms of the classical
$r$-matrix. It provides the rational analogue of the classical
Sklyanin algebra in its original form \cite{Sklyanin}.
The relation between two descriptions ($\eta$-dependent
(\ref{ww003}) and $\eta$-dependent (\ref{ww0081})) comes from both
-- the limit $\eta\to 0$ and from the explicit change of variables:
 \beq\label{ww009}
 \begin{array}{c}
   \displaystyle{
\mathcal S (\eta,\tilde S)=\frac{1}{2}\,{\tilde
L}(\frac{\eta}{2},\tilde S)\,,
 }
 \\ \ \\
  \displaystyle{
 L^\eta\Big(z-\frac{\eta}{2}\,,{\tilde
L}(\frac{\eta}{2},\tilde S)\Big)=\frac{\tr
L^\eta\left(z-\frac{\eta}{2},\tilde S\right)}{\tr\, \tilde
S}\,{\tilde L}(z,\tilde S)\,.
 }
  \end{array}
 \eq
See details in next Section. Both descriptions based on the
quadratic Poisson brackets ((\ref{ww003}) and (\ref{ww0081})) can be
considered as top-like forms of the rational 2-body
Ruijsenaars-Schneider (RS) model \cite{Ruijs1}, while the linear is
the top-like form of the 2-body Calogero-Moser (CM) model
\cite{Calogero}. We will show that  the equation (\ref{ww0071})
describes both - CM and RS models. The difference is in the Poisson
structures and the Hamiltonians.

\vskip3mm

\noindent{\bf Limit to XXX case.} In order to get the standard XXX
$R$-matrices consider the following deformations of (\ref{w001}):
 \beq\label{w0010}
 \begin{array}{c}
  \displaystyle{R^{\hbar,\epsilon}(z)=\epsilon\,R^{\,\epsilon\hbar}(\epsilon
  z)\,,\ \ \ r^{\epsilon}(z)=\epsilon\,r(\epsilon
  z)\,,
  }
 \end{array}
 \eq
i.e.
 \beq\label{w0011}
   \displaystyle{
{r}_{12}^\epsilon(z)=
 \left(\begin{array}{cccc}
1/z & 0 & 0 & 0\\ -z\epsilon^2 & 0 & 1/z & 0\\ -z\epsilon^2 & 1/z & 0 & 0\\
-z^3\epsilon^4 & z\epsilon^2 & z\epsilon^2 & 1/z
 \end{array}
 \right)
 }
 \eq
and similarly for the quantum $R$-matrix. Then
 \beq\label{w0012}
 \begin{array}{c}
  \displaystyle{
  \lim\limits_{\epsilon\rightarrow 0}R^{\hbar,\epsilon}(z)=R^{\hbox{\tiny{XXX}}}(z)=\frac{1}{\hbar}\,
  1\otimes 1+\frac{1}{z}\,P_{12}\,,
  }
  \\
  \displaystyle{
  \lim\limits_{\epsilon\rightarrow
  0}r^{\epsilon}(z)=r^{\hbox{\tiny{XXX}}}(z)=\frac{1}{z}\,P_{12}\,.
  }
 \end{array}
 \eq
The Lax matrices (\ref{ww003}), (\ref{ww007}) and (\ref{ww0081}) are
written in terms of the quantum and classical $R$-matrices. These
Lax operators are the building blocks for more complicated
integrable systems (see below). Then the limit $\epsilon\to 0$
describes transition to the XXX-type models for all the systems
considered in this paper.
The constant parameter $\epsilon$ can be treated as a coupling
constant in integrable tops because the standard XXX case
corresponds to free motion $\dot S=0$.

\vskip3mm

\noindent{\bf Spin chains and Gaudin models.} Having the quadratic
Poisson structure (\ref{ww004}) the classical periodic spin chain
with $n$ sites is naturally defined \cite{FT} via the transfer
matrix
 \beq\label{w0013}
 \begin{array}{c}
  \displaystyle{
  T(z,\mathcal S^1,...,\mathcal S^n)=T(z)=L^{\eta_1}(z-z_1,\mathcal S^1)\,...\,L^{\eta_n}(z-z_n,\mathcal
  S^n)\,,
  }
 \end{array}
 \eq
where $z_k$ are the inhomogeneities parameters. In our case
(\ref{ww003}) it has the quasi-classical form:
 \beq\label{w0014}
 \begin{array}{c}
  \displaystyle{
  T_0(z)=\tr_{1...n}\left(R^{\eta_1}_{01}(z-z_1)\,...\,R^{\eta_n}_{0n}(z-z_n)\,
  ({\mathcal S}^1)_1\,...\,({\mathcal S}^n)_n\right)\,,
  }
 \end{array}
 \eq
where the index $0$ is for the "auxiliary" space of the classical
Lax representation. In the non-relativistic limit $\eta\to 0$ it
gives rise to the Lax operator of the Gaudin model:
 \beq\label{w0015}
 \begin{array}{c}
  \displaystyle{
  L_0^{\hbox{\tiny{G}}}(z)=\sum\limits_{a=1}^n
  \tr_a\left(r_{0a}(z-z_a)S^a\right)\stackrel{(\ref{ww007})}{=}\sum\limits_{a=1}^n
  L_0(z-z_a,S^a)\,.
  }
 \end{array}
 \eq
 To construct the finite chain one needs to have solutions of the
 reflection equations \cite{Skl_refl}. While the $\eta$-independent Lax matrix (\ref{ww0081}) satisfies the standard reflection equation
 (\ref{wy4}),
 the $\eta$-dependent (\ref{ww003}) requires a small
 modification due to (\ref{ww009}):
 \beq\label{w0016}
 \begin{array}{c}
  \displaystyle{
\{L^\eta_1(z)\,, L^\eta_2(w)\}=\frac{1}{2}[ L^\eta_1(z)\,
L^\eta_2(w),r_{12}(z-w)]- }
 \\ \ \\
  \displaystyle{
-\frac{1}{2}\,L^\eta_1\,(z)r_{12}(z+w+\eta)\,L^\eta_2(w)+\frac{1}{2}\,L^\eta_2(w)\,r_{12}(z+w+\eta)\,L^\eta_1(z)\,.
 }
 \end{array}
 \eq
 We consider the Gaudin
 models and spin chains in Section \ref{spin}.

\vskip3mm

\noindent{\bf 1+1 models  and soliton equations.} For the
homogeneous $z_k=0$ spin chain (\ref{w0014}) the continuous limit
leads to the 1+1 field theories, which are integrable in the sense
of the classical inverse scattering method \cite{ZS}. The equations
of motion are generated by (the zero curvature condition) the
Zakharov-Shabat equations \cite{ZS}:
 \beq\label{w0017}
 \begin{array}{c}
  \displaystyle{
  \p_t U-k\p_x V=[U,V]\,,
  }
 \end{array}
 \eq
where $U$ and $V$ are ${\rm gl}_2$-valued functions on the circle
(with the coordinate $x$). They also depend on the spectral
parameter and dynamical fields $S(x)$.
It was shown in \cite{LOZ} that the mechanical (0+1) models
described by non-dynamical $r$-matrix can be generalized to 1+1
field theory (\ref{w0017}) straightforwardly: one should simply use
the same Lax operator (\ref{ww007}):
 \beq\label{w0018}
 \begin{array}{c}
  \displaystyle{
  U^{\hbox{\tiny{LL}}}(z,S(x))=L(z,S(x))=\tr_2(r_{12}(z)S_2(x))\,.
  }
 \end{array}
 \eq
As will be shown in Section \ref{LL} it leads to Landau-Lifshitz
\cite{LL0,SklyaninLL} type equation:
  \beq\label{w0019}
 \begin{array}{c}
  \displaystyle{
\p_tS=\al [S,S_{xx}]+[S,J(S)]\,,
  }
 \end{array}
 \eq
where $S_{xx}=\p_x^2S$, $J(S)$ is the same as in the top case and
$\al$ is a constant. In the light of the Symplectic Hecke
Correspondence \cite{LOZ} this type of the Landau-Lifshitz model is
equivalent to the (rational ${\rm sl}_2$) 1+1 Calogero field theory
\cite{Krich2,LOZ}.

In the same way we consider the 2-poles case
 \beq\label{w0020}
 \begin{array}{c}
  \displaystyle{
  U^{\hbox{\tiny{chiral}}}(z,S(x))=L(z-z_1,S^1(x))+L(z-z_2,S^2(x))\,.
  }
 \end{array}
 \eq
and get the principal chiral model \cite{ZM,Chered,FT,Klimcik} in
the form:
  \beq\label{w6703}
  \left\{
 \begin{array}{l}
  \displaystyle{
\p_tS^1-k\p_xS^1=-2[S^1,L(z_1-z_2,S^2)]\,,
  }
  \\ \ \\
  \displaystyle{
\p_tS^2+k\p_xS^2=-2[S^2,L(z_1-z_2,S^1)]\,.
  }
 \end{array}
 \right.
 \eq
At last, we describe the  1+1 Gaudin models which can be regarded as
the interacting Landau-Lifshitz magnets. See Section \ref{1+1}.

%

\vskip3mm

\noindent{\bf Purpose of the paper}. In this paper we study
relationships between the simplest examples of 2-body systems of
Calogero-Moser model and Ruijsenaars-Schneider and their relations
to integrable tops. We give accurate description for all cases and
show that the models can be described in a similar way. As by
product we notice that top's description allows naturally to deal
with the reflection equation which makes possible to construct
finite chains on a lattice. At last we mention that our description
is adequate for constructing 1+1 field generalizations such as
principal chiral models and interacting Landau-Lifshitz models. The
examples are non-trivial and new. When parameter $\epsilon$ in
(\ref{w0010}) goes to $0$ we come back to the ordinary Gaudin
models.

\vskip5mm

{\small

\noindent {\bf Acknowledgments.} The work was partially supported by
RFBR grants 12-02-00594 (A.L. and M.O.) and 14-01-00860 (A.Z.). The
work of A.L. was also partially supported by AG Laboratory GU-HSE,
RF government grant, ag. 11 11.G34.31.0023. The work of A.Z. was
also partially supported by the D. Zimin's fund "Dynasty" and by the
Program of RAS "Basic Problems of the Nonlinear Dynamics in
Mathematical and Physical Sciences"  $\Pi$19.

}

\newpage

\section{Relativistic Rational top}
\setcounter{equation}{0}

\subsection{Three descriptions}

Here we outline the algebraic structures given in \cite{LOZ7}. They
are universal, and are valid for  ${\rm gl}_N$ and not only for the rational case.

We give three description of the same classical model:

\vskip1mm

\noindent{\bf 1.} In terms of the linear $r$-matrix structure
(\ref{w006}) with the Lax matrix (\ref{w007}). It is gauge
equivalent to the (spin) Calogero-Moser model. It is further used
for constructing the Gaudin models and 1+1 field theories.

\vskip1mm

\noindent{\bf 2.} In terms of the quadratic $r$-matrix structure
(\ref{w008}) and $\eta$-independent Lax matrix (\ref{w0081}). It is
the conventional form of the classical  Sklyanin algebra. It is used
for constructing the classical spin chains.

\vskip1mm

\noindent{\bf 3.} In terms of the quadratic $r$-matrix structure
(\ref{w004}) and $\eta$-dependent Lax matrix (\ref{w003}). It is
gauge equivalent to the quantum $R$-matrix and the (spin)
Ruijsenaars-Schneider model. The first two descriptions can be
obtained from it by the limit $\eta\to 0$, and the second is also
related to it explicitly (see (\ref{w009})-(\ref{w0091})). It can be
also used for constructing the spin chains.

As it was mentioned in the Introduction the relativistic top is
defined by the classical Lax operator
 \beq\label{w003}
  \begin{array}{c}
  \displaystyle{
 { L^\eta}(z,\mathcal S)\equiv L^\eta(z)=\tr_2 \left(R^{\,\eta}_{12}(z) \mathcal S_2\right)\,,\
 \ \mathcal S=\res\limits_{z=0}  L(z)\,,
 }
 \end{array}
 \eq
where $R^{\,\eta}_{12}(z)$ is the quantum $R$-matrix (\ref{w001}),
and the quadratic $r$-matrix structure
 \beq\label{w004}
 \begin{array}{c}
  \displaystyle{
\{L^\eta_1(z)\,, L^\eta_2(w)\}=[ L^\eta_1(z)\,
L^\eta_2(w),r_{12}(z-w)]\,,
 }
 \end{array}
 \eq
where $r_{12}(z-w)$ is the classical $r$-matrix (\ref{w005}).

The non-relativistic limit $\eta\to 0$ gives rise to the linear
$r$-matrix structure
 \beq\label{w006}
 \begin{array}{c}
  \displaystyle{
\{L_1(z)\,, L_2(w)\}=[ L_1(z) + L_2(w),r_{12}(z-w)]\,.
 }
 \end{array}
 \eq
and the Lax operator
 \beq\label{w007}
  \begin{array}{c}
  \displaystyle{
 { L}(z,S)\equiv L(z)=\tr_2 \left(r_{12}(z) S_2\right)\,,\
 \  S=\res\limits_{z=0}  L(z)\,.
 }
 \end{array}
 \eq
The non-relativistic model is bihamiltonian. In the elliptic case it
was observed in \cite{KLO}. The linear $r$-matrix structure
(\ref{w006}) is compatible with the quadratic one (\ref{w004})
 \beq\label{w008}
 \begin{array}{c}
   \displaystyle{
\{\tilde L_1(z)\,, \tilde L_2(w)\}=[\tilde  L_1(z) \tilde
L_2(w),r_{12}(z-w)]\,,
 }
 \end{array}
 \eq
but with the different Lax operator
 \beq\label{w0081}
 \begin{array}{c}
  \displaystyle{
{\tilde L}(z,\tilde S)={\tilde S}_0\, 1_{2\times 2}+L(z,\tilde
S)-\frac{1}{2}\,\tr L(z,\tilde S)\,1_{2\times 2}\,,
  }
  \\ \ \\
  \displaystyle{
\tilde S_0=\tr\,\tilde S\,,\ \ \ \res\limits_{z=0}\tilde L(z,\tilde
S)=\tilde S -\frac{1}{2}\,\tr\,\tilde S\,1_{2\times 2}\,.
  }
 \end{array}
 \eq
The latter is independent of $\eta$ and provides the rational
analogue of the classical Sklyanin algebra in its original form
\cite{Sklyanin}.
It appears that the Lax matrices (\ref{w003}) and (\ref{w0081})
satisfying quadratic $r$-matrix structures are explicitly
related\footnote{The shift by $\eta/2$ is specific for the rational
case. In the elliptic case it is $\eta$. This difference comes from
the normalization $z\to z/N$ for the rational spectral parameter.}:
 \beq\label{w009}
 \begin{array}{c}
  \displaystyle{
 L^\eta\Big(z-\frac{\eta}{2}\,,{\tilde
L}(\frac{\eta}{2},\tilde S)\Big)=\frac{\tr
L^\eta\left(z-\frac{\eta}{2},\tilde S\right)}{\tr\, \tilde
S}\,{\tilde L}(z,\tilde S)
 }
  \end{array}
 \eq
 There is an explicit change of variables
relating $\eta$-dependent and $\eta$-independent descriptions:
 \beq\label{w0091}
 \begin{array}{c}
  \displaystyle{
\mathcal S (\eta,\tilde S)=\frac{1}{2}\,{\tilde
L}(\frac{\eta}{2},\tilde S)\,.
 }
  \end{array}
 \eq
The coefficient $1/2$ in the r.h.s. is not fixed by (\ref{w009}). We
choose this normalization factor in order to have
$\res\limits_{\eta=0}\mathcal S(\eta,\tilde
S)=\res\limits_{z=0}\tilde L(z,\tilde S)$.

Let us remark here that the first description is naturally related
to the (spin) Calogero-Moser (CM) model while the third is related
to the (spin) Ruijsenaars-Schneider (RS) model \cite{Ruijs1}. In the
spinless case relation is given by the explicit change of variables
generated by the special gauge transformations acting on the Lax
operators (see \cite{AASZ} and \cite{LOZ7}). In view of
(\ref{w0091}) the second  description is also related to the RS
model. In the same time, the first and the second descriptions lead
to the same equations of motion (see (\ref{we05}) below). Hence,
{\em the top's equations of motion can be treated as the common
description for the RS and CM models}. We discuss  this point in
Section \ref{RS}.

 It was shown in
\cite{LOZ7} that there is a number of interrelations between Lax
pairs in different descriptions. For example, the expansion of the
$\eta$-dependent Lax operator (\ref{w003}) near $\eta=0$ provides
$M$-operators for all three descriptions:
  \beq\label{we01}
  \begin{array}{c}
  \displaystyle{
L^\eta(z,\mathcal S)=\eta^{-1}\frac{\tr \mathcal  S}2\,1_{2\times
2}-M(z,\mathcal S) +\eta{\mathcal M}(z,\mathcal S)+O(\eta^2)\,.
 }
 \end{array}
 \eq
The coefficient $M(z,S)=-L(z,S)$ is the $M$-operator for
$L^\eta(z,\mathcal S)$:
  \beq\label{we02}
  \begin{array}{c}
  \displaystyle{
\dot L^\eta(z,\mathcal S)=[L^\eta(z,\mathcal S),M(z,\mathcal S)]\,.
 }
 \end{array}
 \eq
 The next term (${\mathcal M}(z,S)$) in (\ref{we01}) is the $M$-operator for
 the Lax matrices (\ref{w007}) and (\ref{w0081})
 (the $M$-operators are the same since the Lax matrices are differed by only
 scalar terms)
  \beq\label{we03}
  \begin{array}{c}
  \displaystyle{
\dot L(z,S)=[L(z,S),\mathcal M(z,S)]\,.
 }
 \\ \ \\
   \displaystyle{
\dot{\tilde L}(z,\tilde S)=[\tilde L(z,\tilde S),\tilde M(z,\tilde
S)]=[\tilde L(z,\tilde S),\mathcal M(z,\tilde S)]\,.
 }
 \end{array}
 \eq
The Lax equations (\ref{we02}) and (\ref{we03}) give rise to
equations of motion of Euler type:
  \beq\label{we04}
  \begin{array}{c}
  \displaystyle{
\dot{\mathcal S}=[\mathcal S, J^\eta(\mathcal S)]
 }
 \end{array}
 \eq
and
 \beq\label{we05}
  \begin{array}{c}
  \displaystyle{
\dot{ S}=[S, J(S)]\,,
 }
 \\ \ \\
   \displaystyle{
\dot{\ti S}=[{\ti S}, J({\ti S})]\,.
 }
 \end{array}
 \eq
 respectively. The inverse inertia tensor $J(S)$ in (\ref{we05}) can be found from
 (\ref{we03}):
 \beq\label{we06}
  \begin{array}{c}
  \displaystyle{
J:\ \  S\ \rightarrow\ J(S)=\mathcal M(0,S)\,,
 }
 \end{array}
 \eq
 while $J^\eta(\mathcal S)$ is of the form\footnote{It
would be interesting to find out its mechanical treatment among the
known integrable examples of the rigid body motion (see reviews
\cite{STS}).}: 
   \beq\label{we07}
  \begin{array}{c}
  \displaystyle{
 J^\eta:\ \ \mathcal S\ \rightarrow\
 J^\eta(\mathcal S)=\tr_2\left(\left(R_{12}^{\eta,(0)}-r_{12}^{(0)}\right)\mathcal
 S_2\right),
   }
   \end{array}
  \eq
where $R_{12}^{\eta,(0)}$ and $r_{12}^{(0)}$ are the coefficients of
the local expansion of (\ref{w001}) and (\ref{w005})  near
$z=0$\footnote{In  the ${\rm gl}_2$ case, which is under
consideration, (\ref{we07}) is simplified since $r_{12}^{(0)}=0$ for
(\ref{w005}).}:
 \beq\label{we08}
 \begin{array}{c}
  \displaystyle{
 R^\hbar_{12}(z)=\sum\limits_{k=-1}^\infty z^k\,R^{\hbar,(k)}_{12}=\frac{1}{z}\,P_{12}+R^{\hbar,(0)}_{12}+z\,R^{\hbar,(1)}_{12}+O(z^2)\,,
 }
 \\
   \displaystyle{
 r_{12}(z)=\frac{1}{z}P_{12}+r_{12}^{(0)}+O(z)\,.
 }
 \end{array}
 \eq
Notice that plugging the change of variables (\ref{w0091}) into
equations of motion (\ref{we04}) we get the Lax equations for the
$\eta$-independent description
   \beq\label{we09}
  \begin{array}{c}
  \displaystyle{
 \p_t{{\tilde
L}(\frac{\eta}{2},\tilde S)}=\frac{1}{2}\,[{\tilde
L}(\frac{\eta}{2},\tilde S), J^\eta( {\tilde
L}(\frac{\eta}{2},\tilde S))]\,,
   }
   \end{array}
  \eq
 where  $\eta/2$ plays the role of the spectral
 parameter (i.e. (\ref{we09}) is identity in $\eta$).
   In this way we get an alternative definition of $M$-operator for
   the $\eta$-independent description:
   \beq\label{we10}
  \begin{array}{c}
  \displaystyle{
\tilde M(z,\tilde S)=\frac{1}{2}\,J^{2z}( {\tilde L}(z,\tilde S))\,.
   }
   \end{array}
  \eq
  Indeed, one can verify that (cf. (\ref{we03}))
   \beq\label{we11}
  \begin{array}{c}
  \displaystyle{
\frac{1}{2}\,J^{2z}( {\tilde L}(z,\tilde S))=\mathcal M(z,\tilde
S)+\frac{1}{2z}\tilde S_0\,1_{2\times 2}\,.
   }
   \end{array}
  \eq

\subsection{Non-relativistic description}

\noindent {\bf Equations of motion.} Consider the Lie coalgebra
${\rm gl}_2^*$ with coordinates $S_{ij}$
 \beq\label{w290}
 \begin{array}{c}
  \displaystyle{
S=\mat{S_{11}}{S_{12}}{S_{21}}{S_{22}}
  }
 \end{array}
 \eq
and Poisson-Lie brackets
  \beq\label{w291}
 \begin{array}{c}
  \displaystyle{
\{S_{ij},S_{kl}\}=\delta_{il}S_{kj}-\delta_{kj}S_{il}\,,\ \
i,j,k,l=1,2\,.
  }
 \end{array}
 \eq
The Casimir functions are defined by
  \beq\label{w292}
 \begin{array}{c}
  \displaystyle{
C_1=\tr S=S_{11}+S_{22}\,,\ \ \ C_2=\frac{1}{2}\tr
S^2=\frac{1}{2}\left(S_{11}^2+S_{22}^2+S_{12}S_{21}\right)\,.
  }
 \end{array}
 \eq
By fixation of $C_{1,2}$ the space ${\rm gl}_2^*$ (\ref{w290})
reduces to a coadjoint orbit of ${\rm GL}_2$ Lie group. Such orbit
is the phase space of the model. The fixation of $C_{1,2}$ does not
change the brackets (\ref{w291}). On a surface
$C_{1,2}=\hbox{const}$ the brackets are non-degenerated.

Consider the Hamiltonian function
  \beq\label{w293}
 \begin{array}{c}
  \displaystyle{
H=-S_{12}(S_{11}-S_{22})\,.
  }
 \end{array}
 \eq
It generates equations of motion
 \beq\label{w312}
 \left\{
 \begin{array}{l}
  \displaystyle{
\dot S_{11}=-\dot S_{22}=-S_{12}(S_{11}-S_{22})\,,}\\
 \displaystyle{ \dot
S_{21}=-2S_{12}S_{21}+(S_{11}-S_{22})^2\,,}
  \\ \displaystyle{\dot S_{12}=2S_{12}^2\,.}
 \end{array}\right.
 \eq
The latter can be written in the top-like form
 \beq\label{w31}
 \begin{array}{c}
  \displaystyle{
\dot S=\{H,S\}=[S,J(S)]\,,
  }
 \end{array}
 \eq
where the inverse inertia tensor $J$ is the following linear
functional on ${\rm gl}_2$:
  \beq\label{w32}
 \begin{array}{c}
  \displaystyle{
J(S)=-\mat{ S_{12}}{0}{ S_{11}-S_{22}}{-S_{12}}
  }
 \end{array}
 \eq
With this notation the Hamiltonian (\ref{w293}) acquires the form of
the Euler top:
  \beq\label{w321}
 \begin{array}{c}
  \displaystyle{
H=\frac{1}{2}\tr \left(S\,J(S)\right)\,.
  }
 \end{array}
 \eq

\vskip3mm

\noindent {\bf Lax pair.} The Lax matrix equals
 \beq\label{w33}
 \begin{array}{c}
  \displaystyle{
L(z,S)=  \frac{1}{z}\left( \begin{array}{cc} S_{11}-z^2S_{12} &
S_{12}
\\ \ \\ 
S_{21}-z^2 (S_{11}-S_{22})-z^4S_{12} & S_{22}+z^2S_{12}
\end{array} \right)
  }
 \end{array}
 \eq
It has the form
 \beq\label{w334}
 \begin{array}{c}
  \displaystyle{
L(z)=\frac{1}{z}L^{(-1)}+zL^{(1)}+z^3L^{(3)}\,,\ \ \ L^{(-1)}:= S\,,
  }
 \end{array}
 \eq
i.e. it is skew-symmetric
 \beq\label{w333}
 \begin{array}{c}
  \displaystyle{
L(z)=-L(-z)
  }
 \end{array}
 \eq
The generating function for the Hamiltonian(s) is given by
 \beq\label{w34}
 \begin{array}{c}
  \displaystyle{
\frac{1}{2}\tr L^2(z)=\frac{1}{z^2}C_2+2H\,,
  }
 \end{array}
 \eq
where $C_2$ is the Casimir function (\ref{w292}) and $H$ is (\ref{w293}).
The Lax equations
 \beq\label{w35}
 \begin{array}{c}
  \displaystyle{
\dot L(z)=\{H,L(z)\}=[L(z),{\mathcal M}(z)]
  }
 \end{array}
 \eq
with
 \beq\label{w36}
 \begin{array}{c}
  \displaystyle{
{\mathcal M}(z)=-\mat{ S_{12}}{0}{
S_{11}-S_{22}+2z^2S_{12}}{-S_{12}}
  }
 \end{array}
 \eq
 reproduce equations of motion (\ref{w312}).
Note that  the linear combination
  \beq\label{w368}
 \begin{array}{c}
  \displaystyle{
\ti{\mathcal M}(z)=\frac{1}{z}L(z)-\mathcal
M(z)=\frac{1}{z^2}S+z^2\mats{0}{0}{S_{12}}{0}\,.
  }
 \end{array}
 \eq
 have a simple form and is equivalent to $-\mathcal M(z)$ in the Lax equations.

Evaluating the residue of the Lax equation (\ref{w35}) at $z=0$ we
get $\dot S=[S,M(0)]$. Therefore, it follows from (\ref{w31}) that
the inverse inertia tensor and $M$-operator are related as
 \beq\label{w363}
 \begin{array}{c}
  \displaystyle{
 J(S)=\mathcal M(0)\,,
  }
 \end{array}
 \eq
where $J(S)$ is defined by (\ref{w32}). Another simple expression
for $J(S)$ follows from the expansion (\ref{w334}). Taking into
account (\ref{w321})
 \beq\label{w364}
 \begin{array}{c}
  \displaystyle{
 J(S)=L^{(1)}\,.
  }
 \end{array}
 \eq
Finally, let us give one more useful formula (that can be verified
directly):
  \beq\label{w365}
 \begin{array}{c}
  \displaystyle{
 L(z,L(z,S))=\frac{1}{z^2}S+2J(S)\,.
  }
 \end{array}
 \eq

 \vskip3mm

\noindent {\bf Classical $r$-matrix} allows us to write all the
Poisson brackets between matrix elements of the Lax matrix in the
form
 \beq\label{w361}
 \begin{array}{c}
  \displaystyle{
\sum\limits_{i,j,k,l} E_{ij}\otimes
E_{kl}\{L_{ij}(z),L_{kl}(w)\}:=\{L_1(z),L_2(w)\}=[L_1(z)+L_2(w),r_{12}(z-w)]\,,
  }
 \end{array}
 \eq
 where in ${\rm gl}_2$ case
$L_1=L\otimes 1=\mats{L_{11}\,1_{2\times 2}}{L_{12}\,1_{2\times
2}}{L_{21}\,1_{2\times 2}}{L_{22}\,1_{2\times 2}}$, $L_2=1\otimes
L=\mats{L}{0_{2\times 2}}{0_{2\times 2}}{L}$. See, for example,
\cite{FT}. In our case the classical $r$-matrix equals
 \beq\label{w37}
   \displaystyle{
{r}_{12}(z)=
 \left(\begin{array}{cccc}
1/z & 0 & 0 & 0\\ -z & 0 & 1/z & 0\\ -z & 1/z & 0 & 0\\ -z^3 & z & z
& 1/z
 \end{array}
 \right)
 }
 \eq
It can be computed from the quantum one (\ref{w001}) by the
classical limit $\lim\limits_{\hbar\to
0}\left(R^\hbar(z)-\hbar^{-1}1\otimes 1\right)$.
 The $r$-matrix
satisfies the classical Yang-Baxter equation
 \beq\label{w371}
 \begin{array}{c}
  \displaystyle{
[r_{12}(z-w),r_{13}(z)]+[r_{12}(z-w),r_{23}(w)]+[r_{13}(z),r_{23}(w)]=0
 }
 \end{array}
 \eq
and have the following properties:
 \beq\label{w372}
 \begin{array}{c}
  \displaystyle{
r_{12}(z)=-r_{21}(-z)\,,\ \ \ \res\limits_{z=0}r_{12}(z)=P_{12}\,,
 }
 \end{array}
 \eq
where $P_{12}=\sum\limits_{i,j=1}^2 E_{ij}\otimes E_{ij}$ is the
 permutation operator.
The Lax matrix and $r$-matrix are simply related:
 \beq\label{w38}
   \displaystyle{
L(z)=\tr_2\left(r_{12}S_2\right)\,.
 }
 \eq
 The latter  relation is easy to check: evaluate the residue of both parts
 of (\ref{w361}) at $w=0$, and then at $z=0$. It gives  the Poisson-Lie brackets
 (\ref{w291}) in the form:
 \beq\label{w382}
   \displaystyle{
 \{S_1,S_2\}=[S_2,P_{12}]\,.
 }
 \eq
Moreover, plugging (\ref{w38}) into (\ref{w361}) with the Poisson
brackets (\ref{w382}), we obtain  the Yang-Baxter equation
(\ref{w371}).

\vskip3mm

\noindent {\bf Limit to free motion.}
In the limit (\ref{w0011})
 \beq\label{w39}
   \displaystyle{
\lim\limits_{\epsilon\rightarrow 0}\left(\epsilon\,
r_{12}(z\epsilon)\right)=\frac{1}{z}\,P_{12}\,,
 }
 \eq
and we get the trivial system $\dot S=0$ with
$L(z)=\frac{1}{z}\,\tr_2\left(P_{12}S_2\right)=\frac{1}{z}\,S$,
$H=0$.

\subsection{Relativistic top: $\eta$-dependent description}

From  the quantum $R$-matrix (\ref{w001}) written as
$R^\hbar_{12}(z)=\sum\limits_{i,j,k.l=1}^N R^{\,\hbar}_{ij,kl}(z)
\,\mathrm E_{ij}\otimes \mathrm E_{kl}$
we obtain the following Lax matrix:
 \beq\label{w397}
   \displaystyle{
L^\eta(z,\mathcal S)=\sum\limits_{i,j,k,l=1}^N
R^{\,\eta}_{ij,kl}(z)\, \mathrm E_{ij}\, \mathcal S_{lk}=
 }
 \eq
 \beq\label{w070}
 \begin{array}{c}
  \displaystyle{
\frac{1}{z}{\mathcal S}_{2\times
2}+\frac{\tr(S)}{\eta}1_{2\times 2} -(z+\eta)
 \left(
 \begin{array}{cc}
{ {\mathcal S}_{12}}&{0}
\\ \ \\
{({\mathcal S}_{11}-{\mathcal S}_{22}) +(\eta^2+z^2+\eta z){\mathcal
S}_{12}}&{ -{\mathcal S}_{12}}
 \end{array}
 \right)
  }
 \end{array}
 \eq
The Poisson brackets are defined by the quadratic $r$-matrix
structure
 \beq\label{w071}
 \begin{array}{c}
  \displaystyle{
\{L^\eta_1(z)\,, L^\eta_2(w)\}=[ L^\eta_1(z)\,
L^\eta_2(w),r_{12}(z-w)]\,,
 }
 \end{array}
 \eq
with the rational $r$-matrix (\ref{w37}). The Poisson brackets are
written as
 \beq\label{w072}
 \begin{array}{c}
  \displaystyle{
{\mathcal A}_{\hbar=0,\eta}^{\hbox{\tiny{Skl}}}:\ \ \ \{{\mathcal
S}_1,{\mathcal S}_2\}=[{\mathcal S}_1{\mathcal
S}_2,r_{12}^{(0)}]+[L^{\eta,(0)}(S)_1\,{\mathcal S}_2,P_{12}]\,,
 }
 \end{array}
 \eq
where we use notations of the expansion
  \beq\label{w0717}
 \begin{array}{c}
  \displaystyle{
{L^\eta}(z)=\tr_2 \left(R^{\,\eta}_{12}(z) {\mathcal
S}_2\right)=\frac{1}{z} {S}+{ L^{\eta,(0)}}(S)+z\,{
 L^{\eta,(1)}}(S)+O(z^2)\,.
 }
 \end{array}
 \eq
 Since $r_{12}^{(0)}=0$, (\ref{w072}) is simplified
 \beq\label{w073}
 \begin{array}{c}
  \displaystyle{
{\mathcal A}_{\hbar=0,\eta}^{\hbox{\tiny{Skl}}}:\ \ \ \{{\mathcal
S}_1,{\mathcal S}_2\}=[J^{\eta}(S)_1\,{\mathcal S}_2,P_{12}]\,,
 }
 \end{array}
 \eq
This gives the Poisson brackets between the components of $\mathcal
S$:
 \beq\label{w074}
 \begin{array}{c}
  \displaystyle{
\{{\mathcal S}_{11},{\mathcal S}_{12}\}=-\eta^{-1}({\mathcal
S}_{11}+{\mathcal S}_{22}){\mathcal S}_{12}+\eta {\mathcal
S}_{12}^2\,,\ \ \ \{{\mathcal S}_{22},{\mathcal
S}_{12}\}=\eta^{-1}({\mathcal S}_{11}+{\mathcal S}_{22}){\mathcal
S}_{12}+\eta {\mathcal S}_{12}^2
 }
 \\ \ \\
  \displaystyle{
\{{\mathcal S}_{11},{\mathcal S}_{21}\}=\eta^{-1}{\mathcal
S}_{21}({\mathcal S}_{11}+{\mathcal S}_{22})+\eta({\mathcal
S}_{11}^2-{\mathcal S}_{11}{\mathcal S}_{22}-{\mathcal
S}_{12}{\mathcal S}_{21})+\eta^3 {\mathcal S}_{11}{\mathcal
S}_{12}\,,
 }
     \\ \ \\
  \displaystyle{
\{{\mathcal S}_{21},{\mathcal S}_{22}\}=\eta^{-1}{\mathcal
S}_{21}({\mathcal S}_{11}+{\mathcal S}_{22})-\eta({\mathcal
S}_{22}^2-{\mathcal S}_{11}{\mathcal S}_{22}-{\mathcal
S}_{12}{\mathcal S}_{21})+\eta^3 {\mathcal S}_{12}{\mathcal
S}_{22}\,,
 }
  \\ \ \\
  \displaystyle{
\{{\mathcal S}_{12},{\mathcal S}_{21}\}=-({\mathcal
S}_{11}+{\mathcal S}_{22})(\eta^{-1}({\mathcal S}_{11}-{\mathcal
S}_{22})+\eta {\mathcal S}_{12})\,,
 }
   \\ \ \\
  \displaystyle{
\{{\mathcal S}_{11},{\mathcal S}_{22}\}=\eta {\mathcal
S}_{12}({\mathcal S}_{11}-{\mathcal S}_{22}+\eta^2 {\mathcal
S}_{12})\,.
 }
 \end{array}
 \eq
They define the Poisson structure on the phase space of the
relativistic top. The equations of motion have form (\ref{we04})
with
 \beq\label{w41}
 \begin{array}{c}
  \displaystyle{
J^\eta(S)=L^{\eta,(0)}=-\mat{\eta {\mathcal S}_{12}}{0}{\eta^3
{\mathcal S}_{12}+\eta ({\mathcal S}_{11}-{\mathcal S}_{22})}{-\eta
{\mathcal S}_{12}} + \frac{{\mathcal S}_{11}+{\mathcal
S}_{22}}{\eta}\,1_{2\times 2}\,.
  }
 \end{array}
 \eq
 Written in components the equation (\ref{we04}) assumes the form:
 \beq\label{w412}
 \begin{array}{c}
  \displaystyle{
\dot{\mathcal S}_{11}=-\eta {\mathcal S}_{12}({\mathcal
S}_{11}-{\mathcal S}_{22}+\eta^2 {\mathcal S}_{12})=-\dot{\mathcal
S}_{22}\,,\ \ \ \dot{\mathcal S}_{12}=2\eta {\mathcal S}_{12}^2\,,
  }
  \\ \ \\
    \displaystyle{
\eta^{-1}\,\dot{\mathcal S}_{21}=({\mathcal S}_{11}-{\mathcal
S}_{22})^2-2{\mathcal S}_{12}{\mathcal S}_{21}+\eta^2 {\mathcal
S}_{11}{\mathcal S}_{12}-\eta^2 {\mathcal S}_{22}{\mathcal
S}_{12}\,.
 }
 \end{array}
 \eq
The determinant of the Lax matrix (\ref{w070}) defines the Casimir
functions:
 \beq\label{w42}
 \begin{array}{c}
  \displaystyle{
\det{ L}^{\eta}(z)=\frac{1}{z^2}{\mathcal
C}_2+(\frac{1}{z\eta}+\frac{1}{\eta^2}){\mathcal C}_1\,,
  }
 \end{array}
 \eq
 \beq\label{w43}
 \begin{array}{c}
  \displaystyle{
{\mathcal C}_2=\det S={\mathcal S}_{11}{\mathcal S}_{22}-{\mathcal
S}_{12}{\mathcal S}_{21}\,,\ \ \ {\mathcal C}_1=({\mathcal
S}_{11}+{\mathcal S}_{22}+\eta^2 {\mathcal S}_{12})^2-4\eta^2
{\mathcal S}_{12}{\mathcal S}_{22}\,.
  }
 \end{array}
 \eq
The $M$-operator for the Lax equations (\ref{we02}) reproducing
equations of motion (\ref{we04}) is obtained via  (\ref{we01}):
 \beq\label{w44}
 \begin{array}{c}
  \displaystyle{
M(z)=  -\frac{1}{z}\left( \begin{array}{cc} {\mathcal
S}_{11}-z^2{\mathcal S}_{12} & {\mathcal S}_{12}
\\ \ \\ 
{\mathcal S}_{21}-z^2 ({\mathcal S}_{11}-{\mathcal
S}_{22})-z^4{\mathcal S}_{12} & {\mathcal S}_{22}+z^2{\mathcal
S}_{12}
\end{array} \right)\,.
  }
 \end{array}
 \eq

\subsection{Relativistic top: $\eta$-independent description}

Consider the following Lax matrix
 \beq\label{w50}
 \begin{array}{c}
  \displaystyle{
{\tilde L}(z,\tilde S)={\tilde S}_0\, 1_{2\times 2}+L(z,\tilde
S)-\frac{1}{2}\,\tr L(z,\tilde S)\,1_{2\times 2}
  }
 \end{array}
 \eq
or
  \beq\label{w503}
 \begin{array}{c}
  \displaystyle{
{\tilde L}(z)=\left( \begin{array}{cc} {\tilde S}_0 & 0
\\ \ \\ 
0 & {\tilde S}_0
\end{array} \right)+\frac{1}{z}\left( \begin{array}{cc} \frac{1}{2}({\tilde S}_{11}-{\tilde S}_{22})-z^2{\tilde S}_{12}
& {\tilde S}_{12}
\\ \ \\ 
{\tilde S}_{21}-z^2 ({\tilde S}_{11}-{\tilde S}_{22})-z^4{\tilde
S}_{12} & \frac{1}{2}({\tilde S}_{22}-{\tilde S}_{11})+z^2{\tilde
S}_{12}
\end{array} \right)\,.
  }
 \end{array}
 \eq
It consists of ${\rm sl}_2$ part of (\ref{w33}) and additional
generator $\tilde S_0$. The Poisson structure is again the quadratic:
 \beq\label{w502}
 \begin{array}{c}
  \displaystyle{
\{{\tilde L}_1(z)\,, {\tilde L}_2(w)\}=[{\tilde L}_1(z)\, {\tilde
L}_2(w),r_{12}(z-w)]\,,
 }
 \end{array}
 \eq
The classical Sklyanin algebra has the form:
 \beq\label{w51}
 \begin{array}{c}
  \displaystyle{
{\mathcal A}_{\hbar=0,\eta=0}^{\hbox{\tiny{Skl}}}:\ \ \ \{{\tilde
S}_1,{\tilde  S}_2\}={\tilde S}_0\,[{\tilde  S}_2,P_{12}] +[{\tilde
S}_1{\tilde  S}_2,r_{12}^{(0)}]+[\tr_3
 (r_{13}^{(0)}{\tilde  S}_3)\,{\tilde  S}_2,P_{12}]\,.
 }
 \end{array}
 \eq
Substituting $r^{(0)}=0$ we get
 \beq\label{w52}
 \begin{array}{c}
  \displaystyle{
{\mathcal A}_{\hbar=0,\eta=0}^{\hbox{\tiny{Skl}}}:\ \ \ \{{\tilde
S}_1,{\tilde  S}_2\}={\tilde S}_0\,[{\tilde  S}_2,P_{12}] \,.
 }
 \end{array}
 \eq
and
  \beq\label{w53}
  \begin{array}{c}
  \displaystyle{
\{{\tilde S}_0, \tilde S\}=\lim\limits_{\eta\rightarrow
0}\frac{[\tilde S,{ J^\eta}(\tilde  S)]}{\eta}=[\tilde  S,J(\tilde
S)]\,,
 }
 \end{array}
 \eq
i.e. the brackets between ${\rm sl}_2$-variables keep the same form
as in (\ref{w291}) but multiplied by ${\tilde S}_0$, while the
brackets between any of ${\rm sl}_2$-variables and ${\tilde S}_0$
are just the corresponding non-relativistic equations of motion
(\ref{w312}):
  \beq\label{w531}
  \begin{array}{c}
  \displaystyle{
\{{\tilde S}_0,  {\tilde S}_{11}\}=-\{{\tilde S}_0,
   {\tilde S}_{22}\}=- {\tilde S}_{12}( {\tilde S}_{11}- {\tilde S}_{22})\,,\ \
   \{{\tilde S}_0,{\tilde S}_{12}\}=2{\tilde S}_{12}^2\,,}
   \\ \ \\
  \displaystyle{
   \{{\tilde S}_0,{\tilde S}_{21}\}=-2{\tilde S}_{12}{\tilde S}_{21}+({\tilde S}_{11}-{\tilde S}_{22})^2\,,
 }
    \\ \ \\
  \displaystyle{
   \{{\tilde S}_{11},{\tilde S}_{12}\}=-{\tilde S}_0{\tilde S}_{12}\,,\ \  \{{\tilde S}_{11},{\tilde S}_{21}\}={\tilde S}_0{\tilde S}_{21}\,,\
   \ \{{\tilde S}_{12},{\tilde S}_{21}\}={\tilde S}_0({\tilde S}_{22}-{\tilde S}_{11})\,.
 }
 \end{array}
 \eq
  In
other words, to get the quadratic algebra we add the Hamiltonian of
the top to the ${\rm sl}_2$ Lie algebra generators.
The Casimir functions are generated by
  \beq\label{w532}
  \begin{array}{c}
  \displaystyle{
\det{\ti L}(z)=\frac{1}{z^2}{\ti C}_2+{\ti C}_0\,,
 }
 \end{array}
 \eq
   \beq\label{w533}
  \begin{array}{c}
  \displaystyle{
{\ti C}_2=-\frac{1}{4}\left({\ti S}_{11}-{\ti S}_{22}\right)^2-{\ti
S}_{12}{\ti S}_{21}\,,\ \ {\ti C}_0={\ti S}_0^2+2{\ti S}_{12}({\ti
S}_{11}-{\ti S}_{12})\stackrel{(\ref{w293})}{=}{\ti
S}_0^2-2H^{\hbox{\tiny{top}}}(\ti S)\,.
 }
 \end{array}
 \eq
The Lax matrix is also the same (up to the scalar component).
Therefore, the Lax pair is the same.

\noindent{\bf Relation to $\eta$-dependent description}.
The Lax matrices in $\eta$-dependent and $\eta$-independent
descriptions $L^\eta(z,S)$ and ${\tilde L}(z,\tilde S)$ are related
as follows (\ref{w009}):
  \beq\label{w58}
  \begin{array}{c}
  \displaystyle{
L^\eta\left(z+\eta_0,{\tilde
L}(-\eta_0,S)\right)=\phi^\eta(z+\eta_0)\,{\tilde L}(z,S)\,,
 }
 \end{array}
 \eq
 where
  \beq\label{w581}
  \begin{array}{c}
  \displaystyle{
\phi^\eta(z)=\frac{\tr L^\eta\left(z,S\right)}{\tr S}
 }
 \end{array}
 \eq
and
  \beq\label{w582}
  \begin{array}{c}
  \displaystyle{
\eta_0=\eta_0(\eta):\ \ \tr L^\eta\left(\eta_0,S\right)=0\,.
 }
 \end{array}
 \eq
 In ${\rm gl}_2$ case (\ref{w070}) we have
  \beq\label{w583}
  \begin{array}{c}
  \displaystyle{
\phi^\eta(z)=\frac{2z+\eta}{z\eta}\,, \ \ \ \eta_0=-\eta/2\,.
 }
 \end{array}
 \eq
The change of variables can be fixed as in (\ref{w0091})
 \beq\label{w00921}
 \begin{array}{c}
  \displaystyle{
\mathcal S (\eta,\tilde S)=\frac{1}{2}\,{\tilde
L}(\frac{\eta}{2},\tilde S)\,.
 }
  \end{array}
 \eq
 These formulae allows us to pass from $\mathcal S$ to $\ti S$
 (\ref{w503})
  \beq\label{w5831}
  \begin{array}{c}
  \displaystyle{
{\ti S}_{0}=\tr{\mathcal S}={\mathcal S}_{11}+{\mathcal S}_{22}\,,\
\ \ {\ti S}_{11}-{\ti S}_{22}=\eta\,({\mathcal S}_{11}-{\mathcal
S}_{22})+\frac{1}{2}\eta^3 {\mathcal S}_{12}\,,
 }
 \\ \ \\
  \displaystyle{
{\ti S}_{12}=\eta\,{\mathcal S}_{12}\,,\ \ \ {\ti
S}_{21}=\eta\,{\mathcal S}_{21}+\frac{1}{4}\eta^3 ({\mathcal
S}_{11}-{\mathcal S}_{22})+\frac{3}{16}\eta^5 {\mathcal S}_{12}\,.
 }
 \end{array}
 \eq
Notice that transition from the $\eta$-dependent Poisson structure
(\ref{w072})-(\ref{w074}) to the $\eta$-independent
(\ref{w52})-(\ref{w531}) can be also performed by taking the limit
$\eta\to 0$. This is due to the structure of the change of variables
(\ref{w00921}), which contains  a simple pole in $\eta$ (at
$\eta=0$).

\subsection{Ruijsenaars-Schneider and Calogero-Moser
models}\label{RS}

The gauge transformations relating the rational Lax operators of the
RS (CM) and the relativistic (non-relativistic) top models can be
found in \cite{LOZ7} (\cite{AASZ}). For the RS model
 \beq\label{w00}
 \begin{array}{c}
  \displaystyle{
{ L}^{\hbox{\tiny{RS}}}(z)=g^{-1}(z)\, g(z+\eta)\,e^{P/c}\ \to \
L^\eta(z)=g(z){ L}^{\hbox{\tiny{RS}}}(z)g^{-1}(z)=
g(z+\eta)\,e^{P/c}\,g^{-1}(z)\,.
 }
 \end{array}
 \eq
In view of (\ref{ww003}) the RS Lax matrix acquires the form
 \beq\label{w000}
 \begin{array}{c}
  \displaystyle{
{ L}^{\hbox{\tiny{RS}}}(z)=\tr_2\left( g_1^{-1}(z)\,
(\res\limits_{z=0}g_1^{-1}(z))\,R_{12}^\eta(z)\, g_1(z)\,
g_2(\eta)\,e^{P_2/c}\right)\,.
 }
 \end{array}
 \eq
Here we discuss the resultant change of variables (or bosonization
formulae) given in ${\rm gl}_N$ case by
 \beq\label{w58411}
 \begin{array}{c}
  \displaystyle{
  {\mathcal S}_{ij}(\bfq,{\bf p})=\sum_{m=1}^{N}\,\frac{({
q}_{m}+\eta)^{\,\varrho(i)} e^{p_{m}/c} }{ \prod\limits_{k\neq
m}^{\,} ({ q}_{m}-{
q}_{k})}\,\,\,(-1)^{\varrho(j)}\,\sigma_{\varrho(j)}(\bfq)\,,
  }
 \end{array}
 \eq
where $\varrho(i)=i-1$ for ${i\leq N-1}$ and $\varrho(N)=N$, while
$\sigma_k$ are the elementary symmetric functions.

In the center of mass frame set $p_2=-p_1=-p$ and $q_2=-q_1=-q$,
i.e. we deal with one pair of canonical coordinates
  \beq\label{w5841}
  \begin{array}{c}
  \displaystyle{
\{p,q\}=1\,.
 }
 \end{array}
 \eq

 \vskip3mm

\noindent{\bf RS model from $\eta$-dependent description.} The
change of variables (\ref{w58411}) with
 $$
 \sigma_0(q)=-\frac{1}{4}(q_1-q_2)^2=-q^2\,,\ \ \sigma_2(q)=1
 $$
gives
  \beq\label{w5842}
  \begin{array}{c}
  \displaystyle{
{\mathcal S}_{11}(p,q)=-\frac{q}{2}\left(e^{p/c}-e^{-p/c}\right)\,,\
\ \ {\mathcal
S}_{12}(p,q)=\frac{1}{2q}\left(e^{p/c}-e^{-p/c}\right)\,,
 }
 \\ \ \\
  \displaystyle{
{\mathcal
S}_{21}(p,q)=-\frac{q}{2}\left(e^{p/c}(q-\eta)^2-e^{-p/c}(q+\eta)^2\right)\,,
 }
 \\ \ \\
  \displaystyle{
{\mathcal
S}_{22}(p,q)=\frac{1}{2q}\left(e^{p/c}(q-\eta)^2-e^{-p/c}(q+\eta)^2\right)\,.
 }
 \end{array}
 \eq
 In particular, it means that the Poisson brackets (\ref{w074}) follows from
 (\ref{w5842}) and (\ref{w5841}).
Notice that having the factor $1/c$ in the exponents one should also
put it (as a common factor) to the r.h.s. of the Sklyanin algebra
(\ref{w073}). To get it from the initial $r$-matrix structure
(\ref{w071}) one can multiply the $r$-matrix by $1/c$.

The relativistic top's Hamiltonian equals
   \beq\label{w5843}
  \begin{array}{c}
  \displaystyle{
\tr{\mathcal S}(p,q)={\mathcal S}_{11}(p,q)+{\mathcal
S}_{22}(p,q)=\eta \frac{\eta-2q}{2q}e^{p/c}-\eta
\frac{\eta+2q}{2q}e^{-p/c}\,.
 }
 \end{array}
 \eq
 The RS Hamiltonian is proportional to $\tr{\mathcal S}(p,q)$. Let us define it as
    \beq\label{w5844}
  \begin{array}{c}
  \displaystyle{
H^{\hbox{\tiny{RS}}}=-\eta^{-1}\,\tr{\mathcal S}(p,q)=
\frac{2q-\eta}{2q}e^{p/c}+ \frac{2q+\eta}{2q}e^{-p/c}\,.
 }
 \end{array}
 \eq
The passage to $\ti S$ variables can be made via (\ref{w5831}).

 \vskip3mm

\noindent{\bf RS model in $\eta$-independent description} comes from
the $\eta$-dependent (\ref{w5842}) one and the change of variables
(\ref{w5831}). Plugging  (\ref{w5842}) into  (\ref{w5831}) we get
    \beq\label{w5845}
  \begin{array}{c}
  \displaystyle{
{\ti S}_{0}= \eta \frac{\eta-2q}{2q}e^{p/c}-\eta
\frac{\eta+2q}{2q}e^{-p/c}\,,
 }
  \\ \ \\
  \displaystyle{
{\ti S}_{11}-{\ti S}_{22}=-\frac{\eta}{4q}\left(
e^{p/c}(2q-\eta)^2-e^{-p/c}(2q+\eta)^2  \right)\,,\ \ \ {\ti
S}_{12}(p,q)=\frac{\eta}{2q}\left(e^{p/c}-e^{-p/c}\right)\,,
 }
 \\ \ \\
   \displaystyle{
 {\ti S}_{21}(p,q)=-\frac{\eta}{32q}\left( e^{p/c}(2q-\eta)^4-e^{-p/c}(2q+\eta)^4 \right)\,.
 }
 \end{array}
 \eq
 The RS Hamiltonian (\ref{w5844}) is obviously has the form
    \beq\label{w5852}
  \begin{array}{c}
  \displaystyle{
H^{\hbox{\tiny{RS}}}=-\eta^{-1}\,{\ti S}_{0}\,.
 }
 \end{array}
 \eq

 \vskip3mm

\noindent{\bf Calogero-Moser model} appears in the non-relativistic
limit
    \beq\label{w5871}
  \begin{array}{c}
  \displaystyle{
\eta:=\nu/c\,,\ \ \ c\to\infty\,.
 }
 \end{array}
 \eq
For the Hamiltonian (\ref{w5844}) we have
    \beq\label{w5872}
  \begin{array}{c}
  \displaystyle{
H^{\hbox{\tiny{RS}}}=2+\frac{2}{c^2}H^{\hbox{\tiny{CM}}}+o(\frac{1}{c^2})\,,
 }
 \end{array}
 \eq
where
    \beq\label{w5873}
  \begin{array}{c}
  \displaystyle{
H^{\hbox{\tiny{CM}}}=\frac{1}2
p^2-\nu\frac{p}{2q}=\frac{1}{2}\left(p-\frac{\nu}{2q}\right)^2-\frac{1}{2}\frac{\nu^2}{(2q)^2}\,.
 }
 \end{array}
 \eq
The conventional form $H^{\hbox{\tiny{CM}}}=\frac{p^2}2
+\frac{\nu^2}{(2q)^2}$ can be obtained by the substitution
$\nu\to\sqrt{-2}\nu$ and the canonical map $p\to p+\frac{\nu}{2q}$.

Consider also the limit (\ref{w5871}) of the residue  matrix
$\mathcal S$ (\ref{w5842}):
    \beq\label{w5874}
  \begin{array}{c}
  \displaystyle{
S=-\frac{1}{2}\lim\limits_{c\to\infty}{c}\,{\mathcal
S}=\mat{\frac{1}{2}p\,q}{-\frac{1}{2}\frac{p}{q}}{\frac{1}{2}(p\,q^3-2\nu
q^2)}{-\frac{1}{2}p\,q+\nu}
 }
 \end{array}
 \eq
The Poisson brackets between the matrix elements of (\ref{w5874})
are the linear Poisson-Lie (\ref{w291}). The eigenvalues of the
matrix $S$ are equal to  $0$ and $\nu$.

Notice that the equations of motion of the relativistic top in the
$\eta$-independent description have the same form as the
non-relativistic equations  (\ref{we05}). Therefore, equations
(\ref{w312})-(\ref{w31}) describe both - the 2-body RS model via the
change of variables (\ref{w5845}) and the 2-body CM model via
(\ref{w5874}).

The obtained formulae for $\mathcal S(p,q)$ and $S(p,q)$ are
particular cases of those obtained in \cite{LOZ7} and \cite{AASZ} at
the level of classical mechanics. The underlying construction is the
Symplectic Hecke Correspondence \cite{LOZ} (see also reviews
\cite{LOZ5,SZ}). In fact, in the quantum counterpart of the
quantum-classical relation (\ref{ww003}) (i.e. from the Sklyanin
algebra point of view) these type of formulae are known from the
original papers \cite{Sklyanin} (in the elliptic case).

\section{Spin chains and Gaudin models}\label{spin}
\setcounter{equation}{0}

\subsection{Gaudin models}

Consider the phase space consisting of $n$ copies of
(\ref{w290})-(\ref{w292}), i.e. the direct product of coadjoint
orbits of ${\rm GL}_2$. It means that we deal with $S^a$, $a=1...n$
and direct sum of the Poisson-Lie brackets
 \beq\label{w3901}
   \displaystyle{
 \{S_1^a,S_2^b\}=\delta^{ab}\,[S_2^a,P_{12}]\,.
 }
 \eq
Fixation of the Casimir functions $C_{1,2}^a$ leaves $2$-dimensional
space for each $S^a$. Hence, the total dimension of the phase space
is equal $2n$. Consider the Lax matrix written in terms of
(\ref{w33}):
 \beq\label{w04}
 \begin{array}{c}
  \displaystyle{
  L^{\hbox{\tiny{G}}}(z)=\sum\limits_{a=1}^n L(z-z_a, S^a)\,.
 }
 \end{array}
 \eq
The Hamiltonians appear by evaluating
 \beq\label{w07}
 \begin{array}{c}
  \displaystyle{
 \frac{1}{2}\tr\left( L^{\hbox{\tiny{G}}}(z) \right)^2=\frac{1}{2}\sum\limits_{a=1}^n \frac{\tr
 \left(S^a\right)^2}{(z-z_a)^2}-\frac{h_a}{z-z_a}+2h_0
 }
 \end{array}
 \eq
The direct computation gives
 \beq\label{w05}
 \begin{array}{c}
  \displaystyle{
  h_a=\sum\limits_{c\neq a}^n h_{a,c}\,,\ \ \
  h_{a,c}=-\tr \left( S^a\,L(z_a-z_c,
 S^c)\right)=-\tr_{12}\left(r_{12}(z_a-z_c) S^a_1  S^c_2\right)\,,
 }
 \end{array}
 \eq
or explicitly
 \beq\label{w06}
 \begin{array}{c}
  \displaystyle{
  h_{a,c}=-\frac{\tr( S^a  S^c)}{z_a-z_c}
  +(z_a-z_c)\left( S_{12}^a( S_{11}^c- S_{22}^c)+ S_{12}^c( S_{11}^a- S_{22}^a)\right)
  +(z_a-z_c)^3\, S_{12}^a S_{12}^c\,.
 }
 \end{array}
 \eq
and
 \beq\label{w08}
 \begin{array}{c}
  \displaystyle{
h_0=\frac{1}{2}\sum\limits_{b,c\,=1}^n\tr\left(S^b\,{ \mathcal
M}(z_b-z_c,S^c)\right)=-\sum\limits_{b,c\,=1}^n
S^b_{12}(S_{11}^c-S_{22}^c) + S_{12}^b S_{12}^c (z_b-z_c)^2=
 }
 \\
  \displaystyle{
=-\sum\limits_{a=1}^n  S^a_{12}(S_{11}^a-S_{22}^a) -\sum\limits_{b>
c}^n  S^b_{12}(S_{11}^c-S_{22}^c) + S^c_{12}(S_{11}^b-S_{22}^b)+
2S_{12}^b S_{12}^c (z_b-z_c)^2\,,
 }
 \end{array}
 \eq
 where $\mathcal M(z,S)$ is the  $M$-operator (\ref{w36}) with properties (\ref{w363}), (\ref{w364}).

 The Hamiltonians (\ref{w05})-(\ref{w08}) generate equations of
 motion
 \beq\label{w091}
 \left\{\begin{array}{l}
  \displaystyle{
 \p_{t_a} S^b=-[S^b,L(z_a-z_b,S^a)]\,,\ \ b\neq a=1,...,n\,,}
\\ \ \\
  \displaystyle{
\p_{t_a} S^a=\sum\limits_{c\neq a}^n\, [S^a,L(z_c-z_a,S^c)]\,,\ \
a=1,...,n
 }
 \end{array}\right.
 \eq
and
 \beq\label{w101}
 \begin{array}{c}
  \displaystyle{
 \p_{t_0}S^a=[S^a,J(S^a)]+\sum\limits_{c\neq a}\,
 [S^a,\mathcal M(z_a-z_c,S^c)]\,,
 }
 \end{array}
 \eq
 where we used (\ref{w363}).
 Equations (\ref{w091}) and (\ref{w101}) have the
 Lax form
 \beq\label{w102}
 \begin{array}{c}
  \displaystyle{
 \p_{t_d} L^{\hbox{\tiny{G}}}(z)=[L^{\hbox{\tiny{G}}}(z),M^{\hbox{\tiny{G}}}_d]\,,\ \ d=0,1,...,n
 }
 \end{array}
 \eq
with
 \beq\label{w09}
 \begin{array}{c}
  \displaystyle{
 M^{\hbox{\tiny{G}}}_a(z)=-L(z-z_a,S^a)\,,\ \ a=1,...,n
 }
 \end{array}
 \eq
and
 \beq\label{w10}
 \begin{array}{c}
  \displaystyle{
 M^{\hbox{\tiny{G}}}_0(z)=\sum\limits_{c=0}^n \mathcal M(z-z_c,S^c)\,,
 }
 \end{array}
 \eq
 where $\mathcal M(z,S)$ is from (\ref{w36}). Expression for $M_d^{\hbox{\tiny{G}}}(z)$ can
 be obtained from
the classical $r$-matrix structure which is the same as in the
 top case (\ref{w361}):
 \beq\label{w11}
 \begin{array}{c}
  \displaystyle{
\{L_1^{\hbox{\tiny{G}}}(z),L_2^{\hbox{\tiny{G}}}(w)\}=[L_1^{\hbox{\tiny{G}}}(z)+L_2^{\hbox{\tiny{G}}}(w),r_{12}(z-w)]
  }
 \end{array}
 \eq
 with the  $r$-matrix (\ref{w37}). The latter holds because the $r$-matrix structure
 is linear as well as the Poisson brackets (\ref{w3901}).

%

 The flows generated by $h_a$ are not independent since
  \beq\label{w061}
 \begin{array}{c}
  \displaystyle{
  \sum\limits_{c=1}^n h_{c}=0\,.
 }
 \end{array}
 \eq
 Put it differently,
  \beq\label{w062}
 \begin{array}{c}
  \displaystyle{
  -\sum\limits_{c=1}^n M^{\hbox{\tiny{G}}}_{c}(z)=L^{\hbox{\tiny{G}}}(z)\,.
 }
 \end{array}
 \eq
The total number of independent integrals of motion equals  $n$
($n-1$ independent $h_c$ and $h_0$). This coincides with the half of
dimension of the phase space. Hence the model is Liouville
integrable.

In the limit (\ref{w0010})-(\ref{w0012})
  \beq\label{w0622}
 \begin{array}{c}
  \displaystyle{
  L^{\hbox{\tiny{G}}}(z)\stackrel{\epsilon\to
  0}{=}\sum\limits_{a=1}^n\frac{S^a}{z-z_a}\,,
 }
 \end{array}
 \eq
 the inverse inertia tensor
$J(S^a)\to 0$ (and $\mathcal M(z,S)\to 0$) and the Hamiltonian $h_0$
(\ref{w08}) become trivial. In the same time the isotropic limit
restores the common ${\rm GL}_2$ symmetry. It compensates the lost
of one Hamiltonian.

Let us mention  that in the light of the Symplectic Hecke
Correspondence \cite{LOZ} the obtained Gaudin model are gauge
equivalent to those considered in \cite{N}.  An explicit relation to
\cite{N} requires some gauge fixation related to additional
reduction by the (global) Cartan subgroup coadjoint action. The
latter action is a common feature of the models with the dynamical
$r$-matrices.




\subsection{Spin chains and reflection equations}

The classical periodic spin chain on $n$ sites is constructed by
introducing the transfer matrix \cite{FT}:
 \beq\label{wy01}
 \begin{array}{c}
  \displaystyle{
T(z)= L^{\eta_1}(S^1, z-z_1)\,...\, L^{\eta_n}(S^n, z-z_n)\,.
 }
 \end{array}
 \eq
The Lax operators which satisfy the quadratic Poisson relation
(\ref{w004}) or (\ref{w008})
 The transfer matrix also satisfies the quadratic Poisson relations.
  Depending on the
 choice of description we have the quasi-classical or purely
 classical expression:
  \beq\label{wy011}
 \begin{array}{c}
  \displaystyle{
  T_0(z)=\tr_{1...n}\left(R^{\eta_1}_{01}(z-z_1)\,...\,R^{\eta_n}_{0n}(z-z_n)\,
  ({\mathcal S}^1)_1\,...\,({\mathcal S}^n)_n\right)
  }
 \end{array}
 \eq
or
   \beq\label{wy012}
 \begin{array}{c}
  \displaystyle{
 \ti T_0(z)=\tr_{1...n}\left(r_{01}(z-z_1)\,...\,r_{0n}(z-z_n)\,
  {\ti S}^1_1\,...\,{\ti S}^n_n\right)\,.
  }
 \end{array}
 \eq
 The classical local Hamiltonian
 appears as follows: set $z_k=0$ and let the Casimir functions (\ref{w43}) or (\ref{w533}) be
 equal for all the sites. Then one should compute the
 transfer-matrix at point $z_0$ given by condition $\det L^k(z_0)=0$
 (it holds simultaneously for all sites due to above requirements).

To get the spin chain on the finite lattice we also need another
Poisson algebra (classical reflection equation) at the boundaries
\cite{Skl_refl}:
 \beq\label{wy4}
 \begin{array}{c}
  \displaystyle{
\{{\tilde L}_1(z)\,, {\tilde L}_2(w)\}= }
 \\ \ \\
  \displaystyle{
\frac{1}{2}[ {\tilde L}_1(z)\, {\tilde L}_2(w),r_{12}(z-w)]-
\frac{1}{2}\,{\tilde L}_1\,(z)r_{12}(z+w)\,{\tilde
L}_2(w)+\frac{1}{2}\,{\tilde L}_2(w)\,r_{12}(z+w)\,{\tilde
L}_1(z)\,,
 }
 \end{array}
 \eq
It appears by reduction from (\ref{w008}) using the constraints
 \beq\label{wy42}
 \begin{array}{c}
  \displaystyle{
\ti L(z,\ti S)\ti L(-z,\ti S)=\det\ti L(z,\ti S)\,1_{2\times 2}\,.
 }
 \end{array}
 \eq
 The Lax matrix (\ref{w503}) satisfies this condition.

One can verify the following statement:

{\em
 The classical $\eta$-independent Lax operator (\ref{w503})
 satisfies the reflection equation (\ref{wy4}). The resultant
 Poisson brackets for the components of $\ti S$ coincide with
 (\ref{w52})-(\ref{w53}).}

It means that we add for the boundaries two Lax operators
$L^\pm(z,S^\pm)$. They are described by the Lax matrices and satisfy
the same classical algebra (\ref{w52})-(\ref{w53}) but through the
reflection equation (\ref{wy4}). The spin chain transfer-matrix
(with dynamical boundaries) is then defined as
 \beq\label{wy5}
 \begin{array}{c}
  \displaystyle{
T(z)={\ti L}(S^+, z)\, {\ti L}(S^1, z-z_1)\,...\, {\ti L}(S^n,
z-z_n)\, {\ti L}(S^-, z)\, {\ti L}(S^n, z-z_n)\,...\, {\ti L}(S^1,
z-z_1)\,.
 }
 \end{array}
 \eq

Remark that at the boundaries we use the same Lax operators
 as inside the chain. In the elliptic case there is an
opportunity to put the extended Lax operators, which give rise to
the inhomogeneous Sklyanin algebra. It is related to the $BC_1$
elliptic model. The corresponding mechanical model has form of the
gyrostat \cite{LOZ2}.

At last, notice that all the consideration can be performed in terms
of the $\eta$-dependent Lax matrix (\ref{w070}). In particular, due
to (\ref{w009}) we have the following form of the reflection
equation for $L^\eta(z)$:
 \beq\label{w5024}
 \begin{array}{c}
  \displaystyle{
\{L^\eta_1(z)\,, L^\eta_2(w)\}=\frac{1}{2}[ L^\eta_1(z)\,
L^\eta_2(w),r_{12}(z-w)]- }
 \\ \ \\
  \displaystyle{
-\frac{1}{2}\,L^\eta_1\,(z)r_{12}(z+w+\eta)\,L^\eta_2(w)+\frac{1}{2}\,L^\eta_2(w)\,r_{12}(z+w+\eta)\,L^\eta_1(z)\,.
 }
 \end{array}
 \eq

\subsection{Canonical variables and many-body interpretation}

The described Gaudin models and spin chains (as well as their XXX
limits) can be rewritten in the canonical coordinates of the 2-body
CM model (\ref{w5874}) or the 2-body RS model (\ref{w5842}),
(\ref{w5845}). It differs from the standard parametrization of ${\rm
sl}_2^*$ which leads to the Garnier type models \cite{Kuznetsov}.

Consider first the Gaudin model. Let all residues $S^a$ are
parameterized by the canonical coordinates $S^a=S^a(p_a,q_a,\nu_a)$
  \beq\label{w821}
 \begin{array}{c}
  \displaystyle{
\{p_a,q_b\}=\delta_{ab}\,,\ \ a,b=1,...,n\,,
  }
 \end{array}
 \eq
where $n$ is the number of the sites (poles in the Gaudin Lax
matrix). Plugging $S^a(p_a,q_a,\nu_a)$ given by (\ref{w5874}) into
the Gaudin Lax matrix (\ref{w04}) we get
  \beq\label{w822}
 \begin{array}{c}
  \displaystyle{
  L^{\hbox{\tiny{G}}}(z)=\sum\limits_{a=1}^n L(z-z_a, S^a(p_a,q_a,\nu_a))\,.
 }
 \end{array}
 \eq
an integrable $n$-particle integrable system depending on $2n$
constants $\nu_a$ and $z_a$. The same, of course, can be done for
the XXX limit (\ref{w0622}). For example,
  \beq\label{w823}
 \begin{array}{c}
  \displaystyle{
  \tr (S^a S^b)=\left(\frac{p_a}{2q_a}(q_b^2-q_a^2)+\nu_a \right)\left(\frac{p_b}{2q_b}(q_a^2-q_b^2)+\nu_b
  \right)\,.
 }
 \end{array}
 \eq
Rewriting in this way the Gaudin Hamiltonians
(\ref{w05})-(\ref{w08}) we get
  \beq\label{w824}
 \begin{array}{c}
  \displaystyle{
  h_a=-\sum\limits_{c\neq a}\frac{1}{z_a-z_c}\left(\frac{p_a}{2q_a}(q_c^2-q_a^2)+\nu_a
  \right)\left(\frac{p_c}{2q_c}(q_a^2-q_c^2)+\nu_c
  \right)+
  }
  \\ \ \\
  \displaystyle{
  +\frac{z_a-z_c}{2}\left(\frac{p_a}{q_a}(p_cq_c-\nu_c)+\frac{p_c}{q_c}(p_aq_a-\nu_a)\right)-(z_a-z_c)^3\frac{p_ap_c}{4q_aq_c}
 }
 \end{array}
 \eq
 and
  \beq\label{w825}
 \begin{array}{c}
  \displaystyle{
  h_0=\sum\limits_{a=1}^n
  \frac{p_a}{2q_a}(p_aq_a-\nu_a)+\frac{1}{2}\sum\limits_{b>c}
  \left(\frac{p_b}{q_b}(p_cq_c-\nu_c)+\frac{p_c}{q_c}(p_bq_b-\nu_b)-(z_b-z_c)^2\frac{p_bp_c}{2q_bq_c}\right)\,.
  }
 \end{array}
 \eq
{\em The first sum in (\ref{w825}) equals the sum of 2-body CM
Hamiltonians $\sum\limits H^{\hbox{\tiny{CM}}}(p_a,q_a,\nu_a)$
(\ref{w5873}). Therefore, this Hamiltonian describes $n$ particles
with masses $m_j\sim\nu_j^2$ in the     central field $\sim
m_j/q_j^2$ with  additional non-trivial interaction.
Notice that this system depends on $2n$ free parameters $\{z_a\}$,
$\{\nu_a\}$. The deformation parameter $\epsilon$ (\ref{w0012}) can
be added as well.}
In the XXX limit the Hamiltonian $h_0$ vanishes, and only the upper
line of (\ref{w824})  survives for $h_a$.

Similar calculations can be made for the spin chain (\ref{wy011}) or
(\ref{wy012}). One can use parametrization in canonical (RS)
variables $\mathcal S^a(p_a,q_a,\eta_a)$ (\ref{w5842}) or $\ti
S^a(p_a,q_a,\eta_a)$ (\ref{w5845})  respectively. For example,
similarly to  (\ref{w823})
  \beq\label{w826}
 \begin{array}{c}
  \displaystyle{
  \tr (\mathcal S^a(p_a,q_a,\eta_a) \mathcal
  S^b(p_b,q_b,\eta_b))=
  }
  \\ \ \\
  \displaystyle{
  \left[\frac{q_a}{2}\left(e^{p_b/c}-e^{-p_b/c}\right)-\frac{1}{2q_a}\left(e^{p_b/c}(q_b-\eta_b)^2-e^{-p_b/c}(q_b+\eta_b)^2\right)\right]\times
  }
    \\ \ \\
  \displaystyle{
  \left[\frac{q_b}{2}\left(e^{p_a/c}-e^{-p_a/c}\right)-\frac{1}{2q_b}\left(e^{p_a/c}(q_a-\eta_a)^2-e^{-p_a/c}(q_a+\eta_a)^2\right)\right]\,.
  }
 \end{array}
 \eq
When $\eta_a=\eta_b=\eta$ this can be used for rewriting the XXX
local Hamiltonian $\sum\limits_{k}\tr(\mathcal S^k \mathcal
S^{k+1})$.

\section{1+1 models}\label{1+1}
\setcounter{equation}{0}
Here we consider the models, which are integrable in the sense of
existing of the Zakharov-Shabat equations \cite{ZS}:
  \beq\label{w60}
 \begin{array}{c}
  \displaystyle{
\p_t U(z)-k\p_x V(z)=[U(z)\,,V(z)]\,,
  }
 \end{array}
 \eq
 where $x$ is a coordinate on the circle.
The dynamical variables become the periodic fields with the Poisson
brackets:
  \beq\label{w672}
 \begin{array}{c}
  \displaystyle{
\{S_{ij}(x),S_{kl}(y)\}=\left(S_{kj}(x)\delta_{il}-S_{il}(x)\delta_{kj}\right)\,\delta(x-y)\,.
  }
 \end{array}
 \eq
We keep notation $S(x)=S$. The procedure of 1+1 generalization of
the models described by non-dynamical $r$-matrices is simple (in
contrast to the case of dynamical $r$-matrices related to many-body
systems, see \cite{LOZ}) -- one should use the same Lax matrix ($U$
matrix) as in the mechanical (top) case. The problem of finding $V$
in general case is more complicated. In the cases under
consideration we will use the ansatz from \cite{SklyaninLL} and its
natural generalization \cite{Z}.

\subsection{Landau-Lifshitz equation}\label{LL}

Set $S_{22}(x)=-S_{11}(x)$ and let $S^2=\lambda^2\,1$, $\p_x\la=0$.
Consider the U-V pair:
 \beq\label{w61}
 \begin{array}{c}
  \displaystyle{
U^{\hbox{\tiny{LL}}}=L(z,S(x))=  \frac{1}{z}\left( \begin{array}{cc}
S_{11}-z^2S_{12} & S_{12}
\\ \ \\ 
S_{21}-2z^2 S_{11}-z^4S_{12} & -S_{11}+z^2S_{12}
\end{array} \right)
  }
 \end{array}
 \eq
and
  \beq\label{w62}
 \begin{array}{c}
  \displaystyle{
V^{\hbox{\tiny{LL}}}=-\frac12(V_1^{\hbox{\tiny{LL}}}+V_2^{\hbox{\tiny{LL}}})
  }
 \end{array}
 \eq

  \beq\label{w63}
 \begin{array}{c}
  \displaystyle{
V_1^{\hbox{\tiny{LL}}}=\frac{1}{z}L(z,S)-2{\mathcal M}(z,S)
=\frac{1}{z^2}\mat{S_{11}}{S_{12}}{S_{21}}{-S_{11}}+
\mat{S_{12}}{0}{2S_{11}+3z^2S_{12}}{-S_{12}}\,,
  }
 \end{array}
 \eq
where $\mathcal M$ is from (\ref{w36}),
  \beq\label{w64}
 \begin{array}{c}
  \displaystyle{
V_2^{\hbox{\tiny{LL}}}=L(z,h)=\frac{1}{z}\left( \begin{array}{cc}
h_{11}-z^2h_{12} & h_{12}
\\ \ \\ 
h_{21}-2z^2 h_{11}-z^4h_{12} & -h_{11}+z^2h_{12}
\end{array} \right)
  }
 \end{array}
 \eq
where the matrix $h$ equals
  \beq\label{w66}
 \begin{array}{c}
  \displaystyle{
h=-\frac{k}{4\la^2}[S,S_x]\,,\ \  S_x=\p_x S\,.
  }
 \end{array}
 \eq
Plugging this U-V pair into (\ref{w60}) we get two equations:
  \beq\label{w65}
 \begin{array}{c}
  \displaystyle{
-k\p_x V_1^{\hbox{\tiny{LL}}}=[L,V_2^{\hbox{\tiny{LL}}}]\,,
  }
  \\ \ \\
  \displaystyle{
\p_t L+\frac{1}{2}k\p_x
V_2^{\hbox{\tiny{LL}}}=-\frac{1}{2}[L,V_1^{\hbox{\tiny{LL}}}]=[L,\mathcal
M]
  }
 \end{array}
 \eq
and, hence
  \beq\label{w652}
 \begin{array}{c}
  \displaystyle{
-k\p_x S=[S,h]\,,
  }
  \\ \ \\
  \displaystyle{
\p_t S+(k/2)\p_x h=[S,J(S)]\,.
  }
 \end{array}
 \eq
Due to the relation $SS_x+S_xS=0$ the first equation can be solved
as given in (\ref{w66}).
 Then the second equation assumes the form:
  \beq\label{w67}
 \begin{array}{c}
  \displaystyle{
\p_tS=\al [S,S_{xx}]+[S,J(S)]
  }
 \end{array}
 \eq
with the constant $\al=k^2/8\la^2$. In components we have
(cf.(\ref{w312})):
 \beq\label{w670}
 \left\{
 \begin{array}{l}
  \displaystyle{
\p_t S_{11}=\al S_{12}\p_x^2 S_{21}-\al S_{21}\p^2_x
S_{12}-2S_{12}S_{11}\,,}\\ \ \\
 \displaystyle{ \p_t
S_{21}=2\al S_{21}\p_x^2 S_{11}-2\al S_{11}\p^2_x
S_{21}-2S_{12}S_{21}+4S_{11}^2\,,}
  \\ \ \\ \displaystyle{\p_t S_{12}=2\al S_{11}\p_x^2 S_{12}-2\al S_{12}\p^2_x
S_{11}+2S_{12}^2\,.}
 \end{array}\right.
 \eq
 The Hamiltonian equals
  \beq\label{w671}
 \begin{array}{c}
  \displaystyle{
H^{\hbox{\tiny{LL}}}=\frac{1}{2}\oint {\rm dx}\,\left(\tr(S_x^2)+
\tr(SJ(S))\right)\,.
  }
 \end{array}
 \eq
The limit (\ref{w0011}) to the continuous Heisenberg model
\cite{Takhtajan} can be performed as in the finite-dimensional case
(see (\ref{w0622}) for $n=1$). Therefore, we obtained an integrable
deformation of the  Heisenberg  model.
Let us mention that close (but different) rational Landau-Lifshitz
equations were found recently in the context of AdS/CFT
correspondence \cite{KY}.

\subsection{Principal chiral model}

To get the (anisotropic) principal chiral model \cite{ZM,Chered,FT}
consider the phase space
  \beq\label{w6700}
 \begin{array}{c}
  \displaystyle{
\{S^a_{ij}(x),S^b_{kl}(y)\}=\delta^{ab}\left(S^a_{kj}(x)\delta_{il}-S^a_{il}(x)\delta_{kj}\right)\,\delta(x-y)\,,
\ \ a,b=1,2\,.
  }
 \end{array}
 \eq
and set
  \beq\label{w6701}
 \begin{array}{c}
  \displaystyle{
L^1=L(z-z_1,S^1(x))\,,\ \ \ L^2=L(z-z_2,S^2(x))
  }
 \end{array}
 \eq
 with $L(z,S)$ (\ref{w33}).
Then the U-V pair
  \beq\label{w6702}
  \left\{
 \begin{array}{l}
  \displaystyle{
U^{\hbox{\tiny{chiral}}}=L^1+L^2\,,
  }
  \\ \ \\
V^{\hbox{\tiny{chiral}}}=L^1-L^2\,.
 \end{array}
 \right.
 \eq
for the Zakharov-Shabat equation (\ref{w60}) gives
  \beq\label{w67031}
  \left\{
 \begin{array}{l}
  \displaystyle{
\p_tS^1-k\p_xS^1=-2[S^1,L(z_1-z_2,S^2)]\,,
  }
  \\ \ \\
  \displaystyle{
\p_tS^2+k\p_xS^2=-2[L(z_2-z_1,S^1),S^2]\,.
  }
 \end{array}
 \right.
 \eq
This is the rational analogue of the anisotropic model
\cite{Chered}. To see its relation to the isotropic one, consider
the deformation (\ref{w0011})
  \beq\label{w6706}
 \begin{array}{c}
  \displaystyle{
L_\epsilon(z_1-z_2,S)=\frac{1}{z_1-z_2}S+\delta_\epsilon L\,,
  }
 \end{array}
 \eq
where
  \beq\label{w6707}
 \begin{array}{c}
  \displaystyle{
\delta_\epsilon
L=-\mat{\epsilon^2(z_1-z_2)S_{12}}{0}{2\epsilon^2(z_1-z_2)S_{11}+\epsilon^4(z_1-z_2)^3S_{12}}{-\epsilon^2(z_1-z_2)S_{12}}
  }
 \end{array}
 \eq
In the isotropic (XXX) limit $\epsilon\to 0$ we find
$L^{1}=S^{1}/(z-z_{1})$, $L^{2}=S^{2}/(z-z_{2})$. Then, by setting
$S^\pm=S^1\pm S^2$ on gets the conventional form of the principal
chiral model:
  \beq\label{w67032}
  \left\{
 \begin{array}{l}
  \displaystyle{
\p_tS^--k\p_xS^+=[S^-,S^+]\,,
  }
  \\ \ \\
  \displaystyle{
\p_tS^+-k\p_xS^-=0\,.
  }
 \end{array}
 \right.
 \eq
It is remarkable that in \cite{Chered} the author obtained the
anisotropic chiral model starting from the 1-site XYZ model, i.e.
from the one pole case instead of the two-poles ansatz
(\ref{w6701}). This can be explained by the passage to the
light-cone coordinates
  \beq\label{w6704}
 \begin{array}{c}
  \displaystyle{
\xi=\frac{kt+x}{2k}\,,\ \ \ \eta=\frac{kt-x}{2k}\,.
  }
 \end{array}
 \eq
Taking into account the  skew-symmetry $L(z)=-L(-z)$, (\ref{w67031})
acquires the form:
  \beq\label{w6705}
  \left\{
 \begin{array}{l}
  \displaystyle{
\p_\eta S^1=-2[S^1,L(z_1-z_2,S^2)]\,,
  }
  \\ \ \\
  \displaystyle{
\p_\xi S^2=-2[S^2,L(z_1-z_2,S^1)]\,,
  }
 \end{array}
 \right.
 \eq
Then, the  reduction $\p_\xi S^2=0$ leads to the equation
$[S^2,L(z_1-z_2,S^1)]=0$. It has a particular solution
$S^2|_{red}=-\frac{1}{2}L(z_1-z_2,S^1)$.
From (\ref{w365})
$L(z_1-z_2,S^2|_{red})=-\frac12\left(\frac{1}{(z_1-z_2)^2}S^1+2J(S^1)\right)$.
Then, from the first equation in (\ref{w6705}) we get the top
equations $\p_\eta S^1=[S^1,J(S^1)]$ (\ref{ww0071}). In this sense
the equations (\ref{w6705}) (and hence (\ref{w67031})) can be also
considered as 1+1 generalization of the top described by the single
pole $z_1=0$ Lax matrix (\ref{ww007}).

\subsection{Interacting Landau-Lifshitz magnets}

An arbitrary number ($n$) of poles in $U$
  \beq\label{w67012}
 \begin{array}{c}
  \displaystyle{
U=\sum\limits_{a=1}^n L(z-z_a,S^a(x))
  }
 \end{array}
 \eq
gives rise to the 1+1 Gaudin type model. It was studied in \cite{Z}.
The elliptic formulae obtained in that paper work for the rational
case under consideration as well. It can be treated as the model of
interacting Landau-Lifshitz magnetics in the same sense as the $t_0$
flow of the Gaudin model (\ref{w101}) looks like interacting
tops\footnote{There is another meaning for the "interacting tops"
\cite{LOSZ4} coming from intermediate cases between the purely
dynamical and purely non-dynamical $R$-matrices.} (\ref{ww0071}).
From the spin chain point of view these type of models arise from
combining each $n$ neighbor sites into one. The quantum and
classical (Poisson) Sklyanin-type algebras underlying discrete
version of 1+1 Gaudin model were described in \cite{CLOZ}.

Equations of motion for the 1+1 Gaudin model are of the form:
  \beq\label{w6709}
  \left\{
 \begin{array}{l}
  \displaystyle{
\p_{t_a}S^a=\p_x h^a+[S^a,J(S^a)]+\sum\limits_{c\neq
a}[h^a,L(z_c-z_a,S^c)]-[V^{\hbox{\tiny{LL}}}_1(z_c-z_a,S^c),S^a]\,,
  }
  \\ \ \\
  \displaystyle{
\p_{t_a}S^b=[S^b,V^{\hbox{\tiny{LL}}}_1(z_b-z_a,S^a)-L(z_a-z_b,h^a)]\,,
\ \ b\neq a\,,
  }
 \end{array}
 \right.
 \eq
where $a,b=1,...,n$, $V^{\hbox{\tiny{LL}}}_1$ is given in
(\ref{w63}) and
  \beq\label{w67091}
 \begin{array}{l}
  \displaystyle{
h^a=\al [S^a,\p_x S^a]+\sum\limits_{c\neq a}L(z_a-z_c,S^c)\,.
  }
 \end{array}
 \eq
When $n=1$ the model coincides with the Landau-Lifshitz one
(\ref{w67}).

     \renewcommand{\refname}{{\normalsize{References}}}


 \begin{small}

 \end{small}

\end{document}